\documentclass[twocolumn]{autart}

\RequirePackage{amsmath, amssymb, graphicx, url,
xspace,subfigure,braket,epstopdf}

\usepackage{graphicx,subfigure}
\usepackage{epsfig} 
\usepackage{mathptmx} 
\usepackage{times} 
\usepackage{amsmath} 
\usepackage{amssymb}  
\usepackage{amsfonts}
\usepackage{euscript}
\usepackage{color}

\begin{document}

\begin{frontmatter}

\thanks[footnoteinfo]{This research has been partially supported by
by the project PRIN 2015 2015PJ28EP.
This paper was
not presented at any IFAC meeting. Corresponding author Gianluigi
Pillonetto Ph. +390498277607.}
\title{Absolute integrability of Mercer kernels\\ is only sufficient for RKHS stability}


\author[First]{Mauro Bisiacco}

\address[First]{Department of Information  Engineering, University of Padova, Padova, Italy (e-mail: bisiacco@dei.unipd.it)}

\author[Second]{Gianluigi Pillonetto}

\address[Second]{Department of Information  Engineering, University of Padova, Padova, Italy (e-mail: giapi@dei.unipd.it)}


\begin{abstract}
Reproducing kernel Hilbert spaces (RKHSs) are
special Hilbert spaces in one-to-one correspondence with positive definite maps called kernels.
They are widely employed in machine learning to reconstruct unknown functions
from sparse and noisy data. In the last two decades,
a subclass known as stable RKHSs 
has been also introduced in the setting of linear system identification.
Stable RKHSs contain only absolutely integrable impulse responses   
over the positive real line. Hence, they can be adopted as hypothesis spaces
to estimate linear, time-invariant and BIBO stable dynamic systems from input-output data. 
Necessary and sufficient conditions for RKHS stability 
are available in the literature and it is known that kernel absolute integrability implies stability.
Working in discrete-time, in a recent work we have proved that 
this latter condition is only sufficient. 
Working in continuous-time, it is the purpose of this note to prove that the same result holds also for Mercer kernels.
\end{abstract}
\maketitle

\end{frontmatter}


\section{Introduction}

\begin{figure}
	\begin{center}
		\begin{tabular}{c}
			{ \includegraphics[scale=0.55]{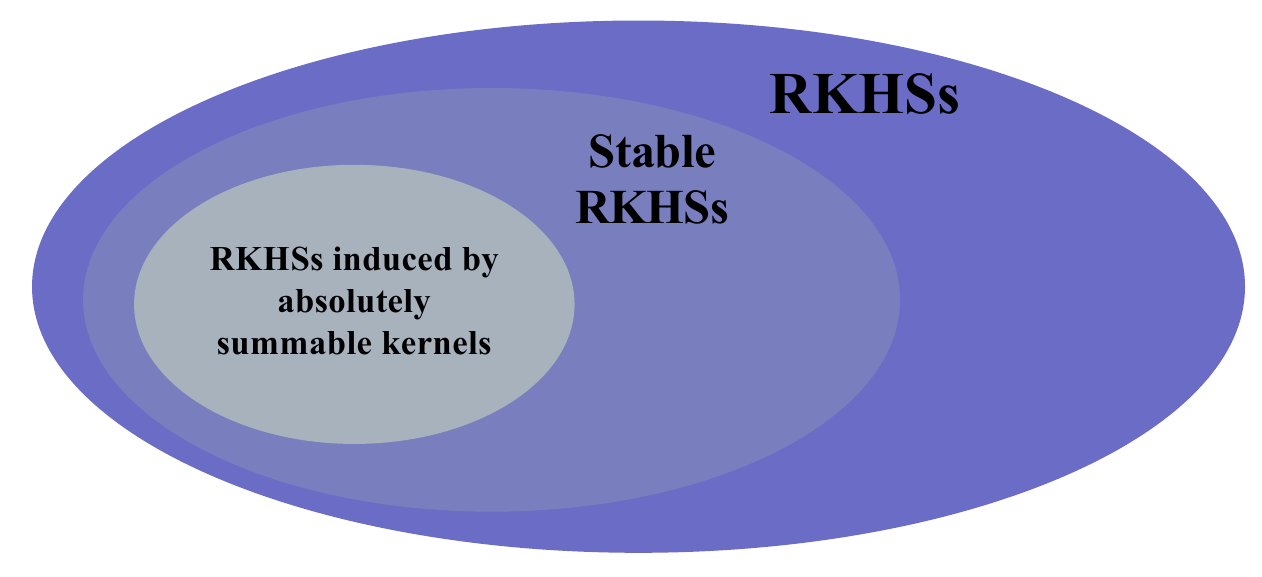}}   
		\end{tabular}
		\caption{Stable Reproducing Kernel Hilbert Spaces (RKHSs) are a subclass of RKHSs containing only absolutely summable real-valued functions over the set of natural numbers or the positive real axis. Hence, they can model time-invariant linear dynamic systems which enjoy BIBO stability. 
		From the necessary and sufficient stability condition reported in \eqref{BIBOH}, it is easy to see
		that kernel absolute summability implies stability but it is hard to understand if also the converse holds.  
		 Extending a result obtained in \cite{AbsSum2020}, this note shows that such condition 
		is not necessary both in discrete- and in continuous-time where Mercer (continuous) kernels are considered.} 
		\label{Fig1}
	\end{center}
\end{figure}

Reproducing kernel Hilbert spaces (RKHSs) are particular Hilbert spaces in one-to-one correspondence
with a positive definite function called kernel. 
They have found important applications in statistics and computer vision starting from the eighties 
\cite{BerteroIEEE,Poggio90,Wahba:90}. Then, they were introduced in the machine learning 
field by Federico Girosi  in \cite{Girosi:1998}.
Their combination with Tikhonov regularization theory \cite{Tikhonov1963,TichonovA:77} leads to   
well known algorithms for supervised learning/function estimation like kernel-ridge regression and support vector machines
\cite{Vapnik98,Scholkopf01b,Suykens2002}.\\
The concept of stable RKHSs for impulse response estimation 
was instead introduced in
\cite{SS2010}. It has been then further developed and applied in many subsequent works like e.g. 
\cite{SS2010,COL12a,SurveyKBsysid,BAHP16}. 
As the name suggests, these spaces contain only absolutely integrable functions and are induced 
by the so called stable kernels \cite{Dinuzzo12,MathFoundStable2020,AbsSum2020}. 
They are now popular since regularization in stable RKHSs
is able to challenge classical approaches to system identification like parametric prediction methods \cite{Ljung:99,Soderstrom}.
Resulting estimators trade-off adherence to experimental data and a penalty term accounting for BIBO stability. 
In this way, among dynamic systems that fit output data in a similar way, the one that is, in some sense, more stable will be chosen \cite{SpringerRegBook2022,PNAS:linsysid2023}.\\
Necessary and sufficient conditions for RKHS stability 
have been obtained in the literature and exploit the kernel operator  \cite{Carmeli,Dinuzzo12}. 
In its continuous-time formulation, for any given kernel $K$, it maps functions $u$ in $y_u$ where
$$
y_u(t)=\int_0^{+\infty} \ K(t,\tau)u(\tau)d\tau, \ t\ge 0.
$$
Then, if $\mathcal{H}$ is the RKHS induced by $K$, one has 
\begin{equation}\label{BIBOH}
\text{BIBO stable $\mathcal{H}$} \ \iff \ y_u \in \mathcal{L}_1 \ \ \forall u \in \mathcal{L}_{\infty} 
\end{equation}
where $\mathcal{L}_{\infty}$ and  $\mathcal{L}_1$ contain, respectively, essentially bounded
and absolutely integrable functions.\\
The result in \eqref{BIBOH} is a full-fledged stability test and one can easily see 
that it is satisfied by any absolutely integrable kernel, i.e. such that
\begin{equation}\label{AbsKernel}
\|K\|_1 := \int_{\mathbb{R}_+^2} \ |K(t,\tau)| dtd\tau < +\infty.
\end{equation}
Indeed, checking \eqref{AbsKernel} is the typical route followed in the literature to guarantee RKHS stability. 
The aim of this note is to understand what happens if \eqref{AbsKernel} does not hold
considering the class of Mercer (continuous) kernels which, in practice, contains all the models adopted in machine learning
and continuous-time system identification.
In particular, the question is: 
\emph{can we claim that the RKHS is unstable if its Mercer kernel is not absolutely summable?}
In \cite{AbsSum2020} a negative answer has been given in the discrete-time setting.
Here, working in continuous-time, we show that the same outcome holds:
also the RKHSs induced by absolutely summable Mercer kernels form a proper subset of the stable RKHSs. 
This result is graphically depicted in Fig. \ref{Fig1} and proved in the remaining part of the paper
through a counterexample. We will build a Mercer kernel which is not absolutely summable but which induces a stable RKHS. 

\section{The counterexample}

To construct the desired counterexample, it is first useful to introduce some notes and new notation. 
Let
\begin{equation}\label{defD}
\mathcal{D}_{\infty} =\Big\{ \ u \in \mathcal{L}_{\infty}: \|u\|_{\infty} \leq 1 \Big\} 
\end{equation}
where $\|\cdot\|_{\infty}$ indicates the norm in $\mathcal{L}_{\infty}$.
A simple application of the closed graph theorem, along the same line e.g. of that described in \cite{CP18}[Lemma 4.1],
shows that kernel stability is equivalent to the continuity
of the kernel operator as a map from  $\mathcal{L}_1$ to $\mathcal{L}_{\infty}$. 
Using $\|K\|_{\infty,1}$ to denote its operator norm, we can reformulate \eqref{BIBOH} as:
\begin{equation}\label{BIBOH2}
\text{BIBO stable $\mathcal{H}$} \ \iff \ \|K\|_{\infty,1} < +\infty 
\end{equation}
where, using $\| \cdot \|_1$  to indicate the norm in $\mathcal{L}_1$ as done in \eqref{AbsKernel},
\begin{equation}\label{Opnorm}
\|K\|_{\infty,1} := \sup_{u \in \mathcal{D}_{\infty}} \ \big\| Ku  \big\|_{1}, 
\end{equation}
with $Ku$ to denote the kernel operator, i.e.
\begin{equation}\label{Ku}
[Ku](\cdot) := \int_0^{+\infty} K(\cdot,x)u(x)dx.
\end{equation}
These notations extend in an obvious way to vectors and matrices. 
If $M$ is e.g. a square matrix of dimension $n$ with components $M_{ij}$, one has 
$$
\| M \|_1 = \sum_{ij} |M_{ij}|
$$ 
and its operator norm is
\begin{equation}\label{Opnorm2}
\|M\|_{\infty,1} = \sup_{v \in \mathbb{R}^n, \| v \|_{\infty}\leq 1} \ \| Mv \big\|_{1}
\end{equation}
with $\| v \|_{\infty}$ given by the maximum of the absolute values of the components of $v$.

\subsection{Continuous approximation of piecewise constant functions}\label{Sub1}

Let
$$
g(x) = \left\{ \begin{array}{cc} 
1 & \mbox{if} \ 0 \le x \le 1 \\
0 &  \mbox{otherwise} 
\end{array}
\right.
$$
with its trapezoidal approximation given by
$$
g_{\epsilon}(x) = \left\{ \begin{array}{cc} 
\frac{1}{2}+\frac{1}{2\epsilon}x & \mbox{for} \ -\epsilon \le x \le \epsilon \\
1 & \mbox{for} \ \epsilon \le x \le 1-\epsilon \\
 \frac{1}{2}+\frac{1}{2\epsilon}(1-x) & \mbox{for} \ 1-\epsilon \le x \le 1+\epsilon \\
 0 & \mbox{elsewhere,}
\end{array}
\right.
$$
where here, and in what follows, $0<\epsilon<\frac{1}{2}$. 
Clearly, one has
\begin{itemize}
\item $g_{\epsilon}(x)$ is continuous in ${\mathbb R}$
\item $g_{\epsilon}(x)$ is zero for $x \le -\frac{1}{2}$ and for $x \ge \frac{3}{2}$
\item $\|g-g_{\epsilon}\|_1=\int_0^{+\infty} \ |g(x)-g_{\epsilon}(x)|dx=\epsilon$.
\end{itemize}
Now, consider a symmetric positive semidefinite matrix $M$ of size $n \times n$ and the associated piecewise constant function ($h,k=1,2,\dots,n$)
$$
\bar{M}(x,y)= 
\left\{ \begin{array}{cc}
M_{hk} & \mbox{for} \ 2h-1 \le x \le 2h, \ 2k-1 \le y \le 2k \\
0 & \mbox{elsewhere}
\end{array}
\right.
$$
One has
$$
\bar{M}(x,y)=\sum_{h,k=1}^n \ M_{hk} \ g(x+1-2h)g(y+1-2k)
$$
and such map can be approximated by the following continuous version 
given by sums of functions with non overlapping supports
$$
\bar{M}_{\epsilon}(x,y) =\sum_{h,k=1}^n \ M_{hk} \ g_{\epsilon}(x+1-2h)g_{\epsilon}(y+1-2k).
$$
The following chain of equalities-inequalities is obtained:
$$
\|\bar{M}-\bar{M}_{\epsilon}\|_1=\int_0^{+\infty}\int_0^{+\infty} \ |\bar{M}(x,y)-\bar{M}_{\epsilon}(x,y)|dxdy=
$$
$$
=\sum_{h,k=1}^n \ |M_{hk}| \ \int_0^{+\infty}\int_0^{+\infty} \ |g_{\epsilon}(x+1-2h)g_{\epsilon}(y+1-2k)-
$$
$$
-g(x+1-2h)g(y+1-2k)|dxdy
$$
(the previous equality holds true since we sum functions with non overlapping supports included in shifted versions of 
the region $[-\frac{1}{2},\frac{3}{2}]^2$)
$$
=\sum_{h,k=1}^n \ |M_{hk}| \  \int_{-\frac{1}{2}}^{\frac{3}{2}}\int_{-\frac{1}{2}}^{\frac{3}{2}} \ |g_{\epsilon}(x)g_{\epsilon}(y)-g(x)g(y)|dxdy
$$
(we are summing functions shifted from each other of compact support)
$$
\le \|M\|_1 \ \int_{-\frac{1}{2}}^{\frac{3}{2}} \ \int_{-\frac{1}{2}}^{\frac{3}{2}} \ |g_{\epsilon}(x)g_{\epsilon}(y)-g(x)g_{\epsilon}(y)|dxdy
$$
$$
+\|M\|_1 \ \int_{-\frac{1}{2}}^{\frac{3}{2}} \  \int_{-\frac{1}{2}}^{\frac{3}{2}} \ |g(x)g_{\epsilon}(y)-g(x)g(y)|dxdy
$$
$$
= \|M\|_1 \ \int_{-\frac{1}{2}}^{\frac{3}{2}} \int_{-\frac{1}{2}}^{\frac{3}{2}} \ \ |g_{\epsilon}(y)| \ |g_{\epsilon}(x)-g(x)|dxdy+\|M\|_1 
$$
$$
\int_{-\frac{1}{2}}^{\frac{3}{2}} \ \int_{-\frac{1}{2}}^{\frac{3}{2}} \ \ |g(x)| \ |g_{\epsilon}(y)-g(y)|dxdy \le \|M\|_1 \ \int_{-\frac{1}{2}}^{\frac{3}{2}} \ \ dy
$$
$$
\int_{-\frac{1}{2}}^{\frac{3}{2}} \ |g_{\epsilon}(x)-g(x)|dx+\|M\|_1 \ \int_{-\frac{1}{2}}^{\frac{3}{2}} \ dx \ \int_{-\frac{1}{2}}^{\frac{3}{2}} \ \ |g_{\epsilon}(y)-g(y)|dy
$$
(both $|g_{\epsilon}(y)|\le 1$ and $|g(x)|\le 1$)
$$
=\|M\|_1 \ \int_{-\frac{1}{2}}^{\frac{3}{2}} \ \ \|g-g_{\epsilon}\|_1dy+\|M\|_1 \ \int_{-\frac{1}{2}}^{\frac{3}{2}} \ \ \|g-g_{\epsilon}\|_1dx=4\|M\|_1\epsilon.
$$
Hence, we have obtained
\begin{equation}
\|\bar{M}-\bar{M}_{\epsilon}\|_1 \le 4\|M\|_1\epsilon.
\label{NORM1}
\end{equation}
Now, letting $M$ be given by $M=VV^T$, with $V$ of size $n \times m$, it holds 
that $M_{hk}=\sum_{r=1}^m \ V_{hr}V_{kr}$, hence
$$
\bar{M}(x,y)=\sum_{h,k=1}^n \ M_{hk} \ g(x+1-2h)g(y+1-2k)=
$$
$$
=\sum_{h,k=1}^n \ \sum_{r=1}^m \ V_{hr}V_{kr} \ g(x+1-2h)g(y+1-2k)=
$$
$$
=\sum_{r=1}^m \ [ \ \sum_{h=1}^n \ V_{hr} \ g(x+1-2h) \ ][ \ \sum_{k=1}^n \ V_{kr} \ g(y+1-2k) \ ]
$$
$$
=:\sum_{r=1}^m \ A_r(x)A_r(y)
$$
and
$$
\bar{M}_{\epsilon}(x,y)=\sum_{h,k=1}^n \ M_{hk} \ g_{\epsilon}(x+1-2h)g_{\epsilon}(y+1-2k)=
$$
$$
=\sum_{h,k=1}^n \ \sum_{r=1}^m \ V_{hr}V_{kr} \ g_{\epsilon}(x+1-2h)g_{\epsilon}(y+1-2k)=
$$
$$
=\sum_{r=1}^m \ [ \ \sum_{h=1}^n \ V_{hr} \ g_{\epsilon}(x+1-2h) \ ][ \ \sum_{k=1}^n \ V_{kr} \ g_{\epsilon}(y+1-2k) \ ]:=
$$
$$
=:\sum_{r=1}^m \ B_r(x)B_r(y)
$$
and this shows that both $\bar{M}$ and $\bar{M}_{\epsilon}$ are positive semidefinite functions.


\subsection{Norms in continuous- and discrete-time}\label{Sub2}

Given the positive semidefinite matrix $M$ of size $n \times n$, consider the functions $\bar{M}(x,y), \ \bar{M}_{\epsilon}(x,y)$ previously introduced and the inputs defined over the interval $0 \le x \le 2n$ with $|u(x)|\le 1$. From integration of piecewise-constant functions,
it is easily seen that
\begin{equation}
\sum_{ij} |M_{ij}|=\|\bar{M}\|_1=\|M\|_1=\int_{\mathbb{R}_+^2} | M(x,y)| dxdy,
\label{NORM2}
\end{equation}
The output $z(x)$ of the kernel operator induced by $\bar{M}(x,y)$ is
$$
z(x)=\int_0^{2n} \ \bar{M}(x,y)u(y)dy=\sum_{k=1}^n \ \int_{2(k-1)}^{2k} \ \bar{M}(x,y)u(y)dy=
$$
$$
=\sum_{k=1}^n \ \int_{2k-1}^{2k} \ \bar{M}(x,y)u(y)dy.
$$
Thus, $z(x)=0$ in any interval of the form $2(h-1) < x < 2h-1, \ h=1,2,\dots,n$, while $z(x)$ is constant and equal to $\bar{z}(h)$ for $h=1,2,\dots,n$ when $2h-1 \le x \le 2h$ again for $h=1,2,\dots,n$, with 
$$
\bar{z}(k)=\sum_{k=1}^n \ M_{hk} \ \int_{2k-1}^{2k} \ u(y)dy:=\sum_{k=1}^n \ M_{hk} \bar{u}(k)
$$
and
$$
\bar{u}(k):=\int_{2k-1}^{2k} \ u(y)dy.
$$
By introducing the vectors $\bar{u},\bar{y}$, with components defined above, it holds that
$$
\bar{z}=M\bar{u}.
$$
Therefore, also $z(x)$ is piecewise-constant and assumes the values $0,\bar{z}(1),0,\bar{z}(2),\dots,0,\bar{z}(n)$. 
This implies
$$
\int_0^{2n} \ |z(x)|dx=\|z\|_1=\|\bar{z}\|_1=\sum_{h=1}^n \ |\bar{z}(h)|.
$$
Letting ${\mathcal D}_{\infty}(I)$ denote the analogous version of ${\mathcal D}_{\infty}$ 
introduced in \eqref{defD} but with the input $u$ restricted to take values on the set $I$,
the above result implies that
$$
\sup_{u \in {\mathcal D}_{\infty}([0,2n])} \ \|\bar{M}u\|_1 = \sup_{\bar{u} \in {\mathcal D}_{\infty}(\{ 1,2,\dots,n \})} \ \|M\bar{u}\|_1 
$$
Indeed, by resorting to inputs $u(x)$ piecewise-constant over intervals of unit length, the corresponding input $\bar{u} \in {\mathcal D}_{\infty}(\{ 1,2,\dots,n \})$ can be any. So, it also holds that
\begin{equation}
\|\bar{M}\|_{\infty,1}=\|M\|_{\infty,1},
\label{NORM3}
\end{equation}
and this, together with \eqref{NORM2}, shows the equivalence between discrete and continuous norms.\\

As far as the ${\mathcal L}_1$-norm is concerned, from \eqref{NORM1} 
we already know that $\|\bar{M}-\bar{M}_{\epsilon}\|_1 \le 4\|M\|_1\epsilon$.
Thus, only the $(\infty,1)-$norm needs to be considered. For any $u \in {\mathcal D}_{\infty}([ \ 0 \ 2n \ ])$, it holds that
$$
z_1(x):=\int_0^{2n} \ \bar{M}(x,y)u(y)dy, \ z_2(x):=\int_0^{2n} \ \bar{M}_{\epsilon}(x,y)u(y)dy \ \Rightarrow 
$$
$$
\Rightarrow \ | \ |z_1(x)|-|z_2(x)| \ | \le |z_1(x)-z_2(x)|
$$
$$
\le \int_0^{2n} \ | \bar{M}(x,y)-\bar{M}_{\epsilon}(x,y) | dy.
$$
So, it follows that
$$
| \ \int_0^{2n} \ |z_1(x)|dx \ - \int_0^{2n} \ |z_2(x)|dx \ | 
$$
$$
\le \int_0^{2n} \ \int_0^{2n} \ | \bar{M}(x,y)-\bar{M}_{\epsilon}(x,y) |=\|\bar{M}-\bar{M}_{\epsilon}  \|_1
$$
which implies
$$
\int_0^{2n} \ |z_1(x)|dx \le \int_0^{2n} \ |z_2(x)|dx + \|\bar{M}-\bar{M}_{\epsilon}\|_1 \ \mbox{and}
$$
$$
\int_0^{2n} \ |z_2(x)|dx \le \int_0^{2n} \ |z_1(x)|dx + \|\bar{M}-\bar{M}_{\epsilon}\|_1.
$$
Taking the superior w.r.t. $u \in {\mathcal D}_{\infty}([0,2n])$, we obtain
$$
\|\bar{M}\|_{\infty,1} \le \|\bar{M}_{\epsilon}\|_{\infty,1}+\|\bar{M}-\bar{M}_{\epsilon}\|_1 \ \mbox{and}
$$
$$
\|\bar{M}_{\epsilon}\|_{\infty,1} \le \|\bar{M}\|_{\infty,1} + \|\bar{M}-\bar{M}_{\epsilon}\|_1
$$
and this implies
$$
| \ \|\bar{M}\|_{\infty,1} - \|\bar{M}_{\epsilon}\|_{\infty,1} \ | \le \|\bar{M}-\bar{M}_{\epsilon}\|_1 \le 4\|M\|_1\epsilon
$$
and, from \eqref{NORM1}, also
$$
| \ \|\bar{M}\|_1 - \|\bar{M}_{\epsilon}\|_1 \ | \le \|\bar{M}-\bar{M}_{\epsilon}\|_1 \le 4\|M\|_1\epsilon.
$$
Summarizing, we have proved in this subsection that 
\begin{equation}
\begin{aligned}
\|\bar{M}\|_{\infty,1}&=\|M\|_{\infty,1}, \ \|\bar{M}\|_1=\|M\|_1, \\
\label{RESULT}
\big|  \|M\|_{\infty,1}& - \|\bar{M}_{\epsilon}\|_{\infty,1} \big| \le 4\|M\|_1\epsilon,\\
\big|  \|M\|_1 &- \|\bar{M}_{\epsilon}\|_1 \big| \le 4\|M\|_1\epsilon.
\end{aligned}
\end{equation}


\subsection{The continuous-time counterexample}\label{Sub3}

Recall that in \cite{AbsSum2020} we built a stable kernel in discrete-time which is not absolutely summable 
exploiting sequence of matrices $M^{(h)}, h=1,2,\dots$,
whose dimensions increase with $h$, such that
$$
\|M^{(h)}\|_1=\frac{1}{h}, \ \|M^{(h)}\|_{\infty,1}\le \frac{1}{h^2}.
$$
These matrices were then used to build a block-diagonal structure. 
For the Mercer kernels in continuous-time here considered, the following workflow is instead adopted:
\begin{itemize}
\item $\bar{M}_{\epsilon_h}^{(h)}(x,y)$ is defined as the approximation of $\bar{M}^{(h)}(x,y)$ and built starting from $M^{(h)}$ according to what done in Section \ref{Sub1}, with $\epsilon_h:=\frac{1}{3h}$ and therefore $0<\epsilon_h<\frac{1}{2}$ for any $h = 1,2,\dots$;
\item a block-diagonal structure is defined by using the $\bar{M}_{\epsilon_h}^{(h)}$ in such a way 
that the supports of such building blocks are all mutually disjoint. Their square supports are placed along the diagonal,  mimicking the discrete-time construction;
\item we define $K(x,y)$ as the corresponding kernel whose continuity, symmetry and positive semidefiniteness is guaranteed by the arguments developed in the previous subsections.
\end{itemize}
Now, it comes from \eqref{RESULT} that
$$
\|K\|_1=\sum_{h=1}^{+\infty} \ \|\bar{M}_{\epsilon_h}^{(h)}\|_1 \ge \sum_{h=1}^{+\infty} \ ( \  \|M^{(h)}\|_1-4\|M^{(h)}\|_1\epsilon_h \ ) 
$$
$$
\ge \sum_{h=1}^{+\infty} \ \frac{1}{h} \ - \ \sum_{h=1}^{+\infty} \ \frac{4}{3h^2} \ \rightarrow \ +\infty \ \Rightarrow \ \|K\|_1=+\infty
$$
and
$$
\|K\|_{\infty,1}=\sum_{h=1}^{+\infty} \ \|\bar{M}_{\epsilon_h}^{(h)}\|_{\infty,1} \le \sum_{h=1}^{+\infty} \ ( \  \|M^{(h)}\|_{\infty,1}+4\|M^{(h)}\|_1\epsilon_h \ ) 
$$
$$
\le \sum_{h=1}^{+\infty} \ \frac{7}{3h^2} < +\infty. 
$$
So, one has 
\begin{equation}
\|K\|_{\infty,1}<+\infty, \ \|K\|_1=+\infty,
\label{URK}
\end{equation}
which proves that $K$ is a stable but not absolutely summable Mercer kernel. 
This concludes the proof.

\section{Conclusions}

As also illustrated in Fig. \ref{Fig1}, the necessary and sufficient condition for RKHS stability
reported in \eqref{BIBOH} does not imply kernel absolute integrability.
This result, known in discrete-time, has been here extended to the continuous-time domain
considering the class of Mercer kernels.
The counterexample here reported proves this fact and, in some sense, 
underlines the complexity and richness of these spaces,
providing further insights about their nature. 


\end{document}